# 1.71 Tb/s Single-Channel and 56.51 Tb/s DWDM Transmission over 96.5 km Field-Deployed SSMF

Fabio Pittalà, *Member*, *IEEE*, Ralf-Peter Braun, Georg Böcherer, *Member*, *IEEE*, Patrick Schulte, *Member*, *IEEE*, Maximilian Schaedler, *Student Member*, *IEEE*, Stefano Bettelli, Stefano Calabrò, *Member*, *IEEE*, Maxim Kuschnerov, Andreas Gladisch, Fritz-Joachim Westphal, Changsong Xie, *Member*, *IEEE*, Rongfu Chen, Qibing Wang, Bofang Zheng

***Abstract* — We report an industry leading optical dense wavelength division multiplexing (DWDM) field trial with line rates per channel ≥1.66 Tb/s using 130 GBaud dual-polarization probabilistic constellation shaping 256-ary quadrature amplitude modulation (DP-PCS256QAM) in a high capacity data center interconnect (DCI) scenario. This research trial was performed on 96.5 km of field-deployed standard single mode G.652 fiber infrastructure of Deutsche Telekom in Germany employing Erbium-doped fiber amplifier (EDFA)-only amplification. A total of 34 channels were transmitted with 150 GHz spacing for a total fiber capacity of 56.51 Tb/s and a spectral efficiency higher than 11bit/s/Hz. In the single-channel transmission scenario 1.71 Tb/s was achieved over the same link. In addition, we successfully demonstrate record net bitrates of 1.88 Tb/s in back-to-back (B2B) using 130 GBaud DP-PCS400QAM.**

***Index Terms* — Coherent communications, Fiber optics components, Optical fiber communication.**

## I. Introduction

Exponential internet traffic growth continues with emerging bandwidth-hungry applications such as 5G, video, artificial intelligence and virtual/augmented reality. In addition, the Covid health crisis has led to a rapid expansion of our digital activity, which will continue to grow beyond the pandemic. In order to ensure sufficient connectivity, constant evolution in research and standardization of high-capacity optical transmission systems is required.

Recent records for single-carrier experiments with net bitrates exceeding 1 Tb/s are summarized in Fig. 1. Filled markers show experiments using one digital-to-analog converter (DAC) per dimension, while non-filled markers show data for transmitters using multiple DACs per dimension. Back-to-back (B2B) experiments are shown in grey. The high-speed DAC is the key component to generate high-bitrate and high-symbol rate signals. The sampling rates of DACs have reached 128 GSa/s for technologies based on Silicon-Germanium (SiGe) [1, 2, 5, 6] and 120 GSa/s for complementary metal-oxide semi-conductor (CMOS) [3]. Using a single DAC per dimension, net

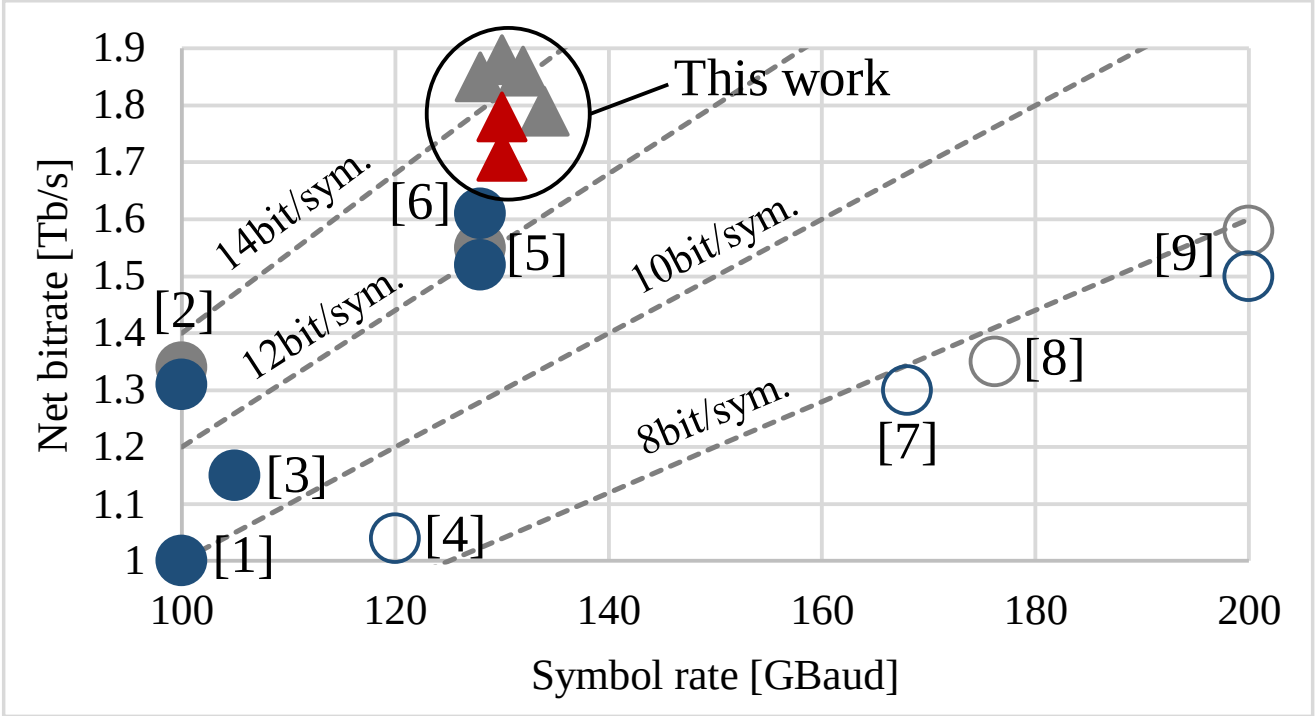

Fig. 1. Net bitrates vs symbol rate for recent record experiments exceeding 1 Tb/s. Filled markers refer to experiments using one DAC per dimension, while non-filled markers show data for transmitters using multiple DACs per dimension. Grey markers indicate B2B demonstrations. Dotted lines show thresholds of constant net spectral efficiencies for polarization multiplexed systems. Triangle markers show the net bitrates achieved in this work.

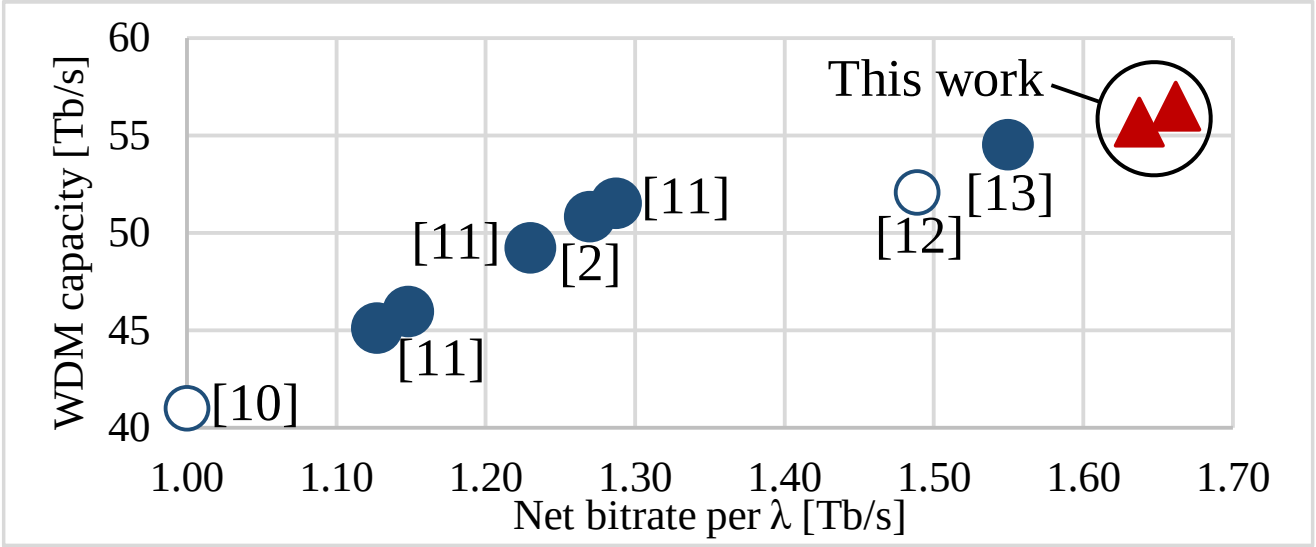

Fig. 2. Reported DWDM capacity vs per carrier net bitrate. Filled markers refer to experiments over field-deployed fiber, while non-filled markers show data for experiments using test fiber in laboratory.

Fabio Pittalà (fabio.pittala@huawei.com), Georg Böcherer, Patrick Schulte, Maximilian Schaedler, Stefano Bettelli, Stefano Calabrò, Maxim Kuschnerov, Changsong Xie are with Huawei Technologies Düsseldorf GmbH, Munich Research Center, Riesstrasse 25, Munich, 80992, Germany (e-mail: fabio.pittala@huawei.com).

Ralf-Peter Braun, Andreas Gladisch and Fritz-Joachim Westphal are with Deutsche Telekom, S&TI, TA&I, Emerging Technologies, Winterfeldtstrasse 21, Berlin 10587, Germany.

Rongfu Chen is with Huawei Technologies Deutschland GmbH, Innovation Center, Hansaallee 205, Duesseldorf, 40549, Germany.

Qibing Wang and Bofang Zheng are with Huawei Technologies Co. Ltd., B&P Laboratory, Shenzhen 518129, China.

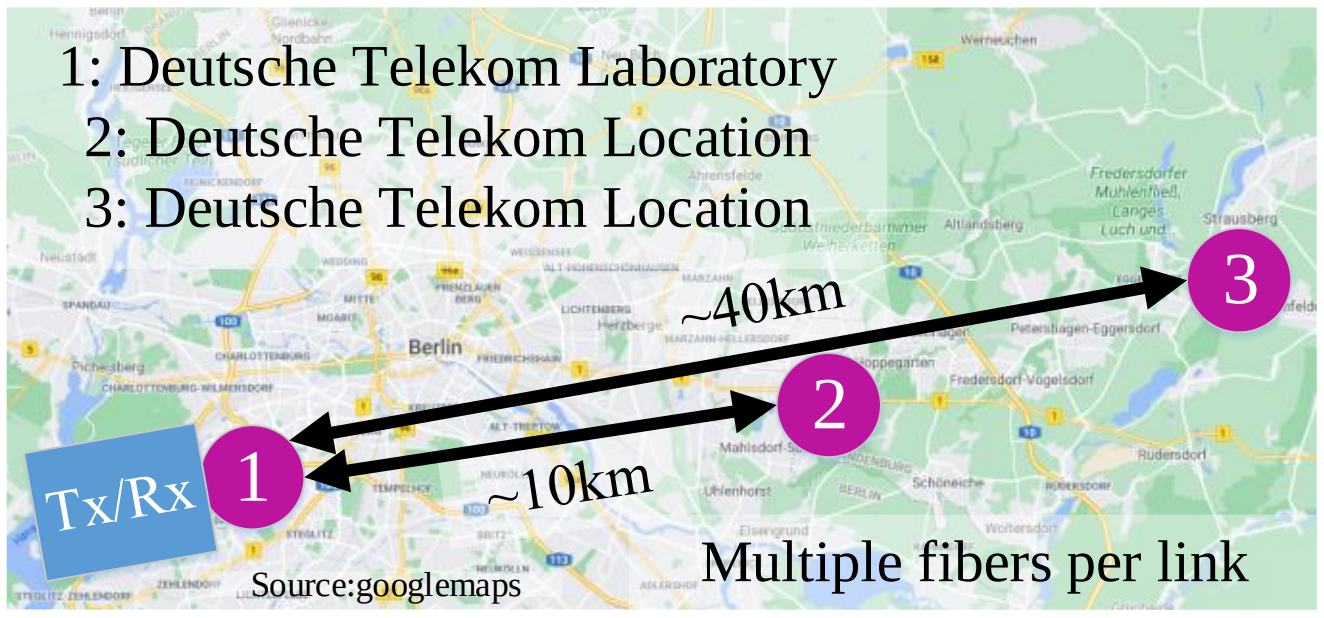

Fig. 3 Field-deployed SSMF infrastructure of Deutsche Telekom used for transmission experiments, multiple fibers per link available and longest configured link reaching 183.9 km.

bitrates up to 1.61 Tb/s have been reported using 128 GSa/s SiGe DACs [6], while the highest reported net bitrate with 120 GSa/s CMOS DACs is 1.15 Tb/s [3]. Using multiple DACs per dimension allows to achieve higher symbol rates, however bandwidth multiplexing techniques affect the signal integrity limiting the achievable information rate (AIR) [4, 7-9]. Recent dense wavelength division multiplexing (DWDM) aggregate capacity records are shown in Fig. 2. In these experiments, the DWDM grid ranges from 100 GHz to 150 GHz and the symbol rates from 96 GBaud up to 130 GBaud. Triangle markers in Fig. 1 and in Fig. 2 show the net bitrates achieved in this work.

We report record symbol rates up to 134 GBaud using off-the-shelf hardware components and in particular using only a single DAC per dimension. We achieved net bitrates up to 1.88 Tb/s for single-carrier optical B2B, 1.78 Tb/s and 1.71 Tb/s, respectively, after transmission over 61.3 km and 96.5 km of field-deployed ITU-T G.652 standard single-mode fiber (SSMF) belonging to the infrastructure of Deutsche Telekom in Germany employing EDFA-only amplification at the Tx/Rx of location ① in Fig. 3. In contrast to all previously reported records, it should be noted that the transmitter used in this paper does not use polarization division multiplexing emulation, but generation of the two polarizations is obtained using four independent lines of components (DAC, driver, modulator) reflecting the actual implementation of a product. In addition, we report the first DWDM transmission experiment over 96.5 km field-deployed SSMF link with a per-carrier net bitrate exceeding 1.6 Tb/s on each of the 34 DWDM channels and yielding to a total capacity of 56.51 Tb/s in C-band with a spectral efficiency of 11.08 bits/s/Hz. To the best of our knowledge, this is the highest single-channel and C-band aggregate net bitrate ever trans-mitted on a field-deployed fiber network of a service provider.

The experimental setup is described in section II, while the main findings are reported in section III. Our conclusions are reported in section IV.

## II. Transmission Setup

The system setup is shown in Fig. 4. The data signal consisting of four real components is generated by four SiGe DAC application-specific integrated circuits (DAC-ASICs) having a typical bandwidth of 65 GHz (with sin(x)/x roll-off mathematically compensated), effective number of bits (ENOB) of 5 bit for the same bandwidth, and a typical sampling rate of up to 128 GSa/s [14]. With respect to the DAC employed in [5, 6], this device uses a novel type of package for the DAC-ASIC and a built-in amplifier delivering high speed and quality output signals with a smooth frequency roll-off. In this experiment we have increased the sampling rate of the DACs up to 134 GSa/s (over the limit indicated in the data sheet of this device) by supplying a reference clock up to 67 GHz. To keep the system stable (synchronized over time with the input reference clock) and to reduce thermal noise in the DAC-ASICs a careful temperature control was performed by using an external air cooling system. Based on the prototype version of this AWG and time interleaving two 128 GSa/s outputs a record symbol rate demonstrating 220 GBaud signal generation has been recently published [15]. The electrical outputs of the DACs are connected to four single-ended SiGe RF amplifiers with 72 GHz 3 dB-bandwidth and 11 dB gain driving two electro-optic GaAs IQ modulators having 6 dB-bandwidth exceeding 50 GHz. Fig. 5 shows the frequency transfer function (without compensating the sin(x)/x roll-off) for the four DAC-ASICs, when the peaking of the built-in amplifier is set to its maximum value, without and with the 72 GHz driver amplifiers. The output optical signal of a tunable external cavity laser (ECL) with <100 kHz linewidth is split in a polarization maintaining (PM) splitter and amplified by two PM Erbium-doped fiber amplifiers (EDFAs) before feeding the two IQ modulators with 18 dBm optical power. The dual-polarization (DP) signal is obtained by recombining the output of the IQ modulators using a polarization beam combiner (PBC). The DP signal is amplified by an EDFA before applying linear pre-emphasis by a waveshaper that, with 8 dB pre-emphasis, flattens the power spectral density of the modulated signal. The transceiver can be tuned over 5.36 THz from 191.1 THz to 196.46 THz. The optical power of the modulated signal at the output of the waveshaper can reach up to 12 dBm.

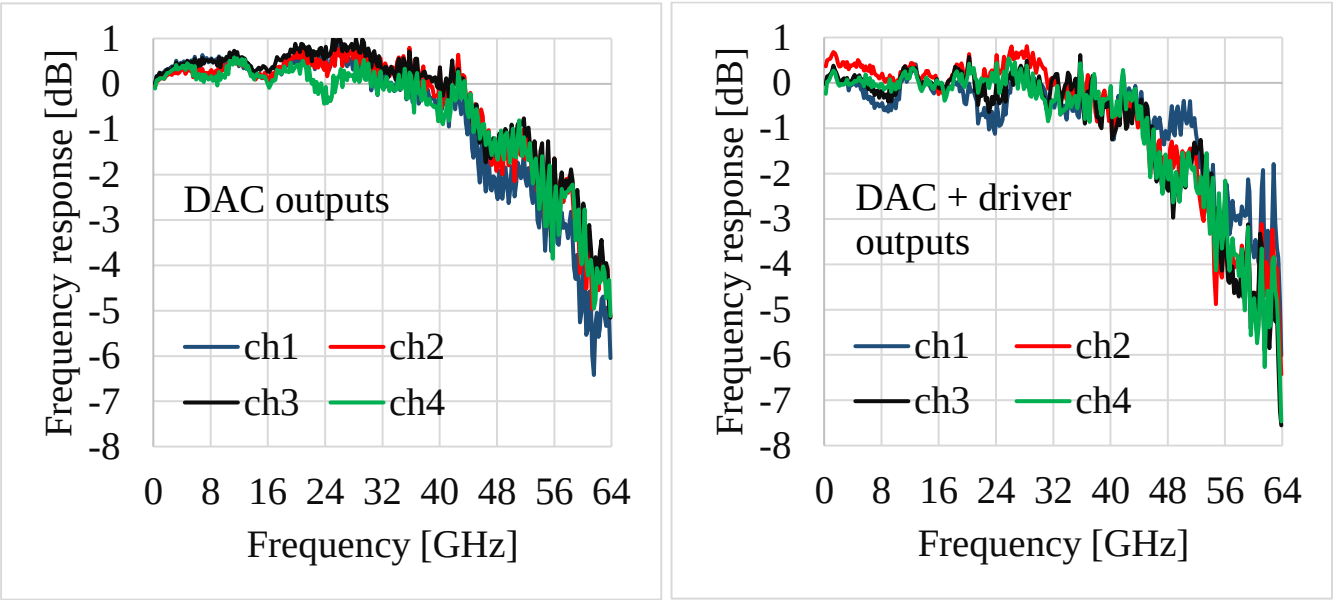

Fig. 5 Frequency transfer function of the four SiGe DAC-ASICs (without sin(x)/x roll-off compensation), setting the peaking of the built-in amplifier to its maximum value, without (left) and with (right) the 72 GHz driver

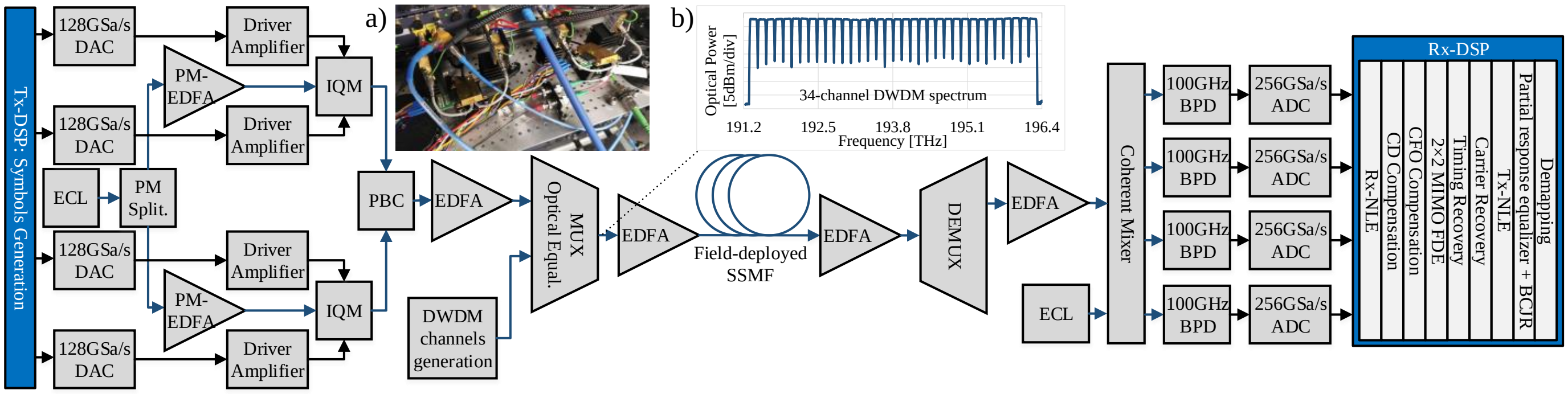

Fig. 4. Schematic of the experimental setup, insets: a) picture of transmitter front-end, b) 34-channel DWDM transmitted spectrum.

The optical links are part of the Deutsche Telekom fiber infrastructure in Germany, as shown in Fig. 3. A set of single-carrier experiments is based on multiple spans of 61.3 km SSMF each having ~19.5 dB attenuation, while the DWDM experiment is based on a single span of 96.5 km of SSMF exhibiting 23 dB attenuation. Amplification is performed using EDFAs only. In B2B and single-channel configurations no booster EDFA is required and after fiber transmission, the signal is pre-amplified before a tunable bandpass filter set to 150 GHz to reduce amplified spontaneous emission noise at the receiver. To emulate a DWDM system, an ASE bandwidth loading method [16] is applied assuming a 150 GHz-spaced 34-channel DWDM from 191.225 THz to 196.325 THz. In the DWDM scenario, optical equalization is done in the MUX followed by a booster amplifier and after selecting the channel under test with the DEMUX, an additional EDFA is used to keep the optical power at the input of the receiver at 7 dBm.

The receiver consists of a coherent mixer and four 100 GHz balanced photodetectors (BPDs), with 0.45A/W responsitivity, connected to a 256 GSa/s 80 GHz oscilloscope. Another ECL with linewidth <100 kHz is used as local oscillator. The receiver digital signal processing (DSP) makes use of advanced and fully adaptive nonlinear component equalizers, targeting imperfections such as bandwidth limitations, frequency dependent I/Q imbalance and skew, phase ripple, I/Q crosstalk and high-order nonlinearities at transmitter and receiver. At the receiver, a first digital Volterra equalizer (Rx-NLE in Fig. 4) addresses the imperfections of the receiver components, i.e. optical-electronic frontend and analog-to-digital converter (ADC). After channel equalization and demodulation (including carrier phase recovery), another Volterra equalizer (Tx-NLE in Fig. 4) compensates for the residual imperfections of the transmitter. Since the transponder imperfections arise mostly in the electrical domain, where the four tributaries are independently processed, also the equalizers operate on the real tributaries rather than on the complex baseband signal. Finally, partial-response equalization (PREQ) with impulse response 1+αD is implemented to whiten the noise, followed by a complex-valued BCJR algorithm with one memory tap used for sequence detection, similarly to [17, 18].

## III. Measurement Results

Measurements are performed using a family of probabilistic constellation shaping (PCS) formats of variable entropy (H) obtained from high-order square quadrature-amplitude modulation (QAM) formats: 256QAM, 324QAM, 400QAM, 484QAM and 576QAM, compatible with probabilistic amplitude shaping (PAS) [19]. In B2B, for each base constellation and symbol rate (128-134 GBaud) we varied the alphabet entropy using Maxwell-Boltzmann distributions. Then we selected the configuration resulting in the maximum AIR: 256QAM with H=7.9 bit/symbol, 324QAM with H=8.14 bit/symbol, 400QAM with H=8.44 bit/symbol, 484QAM with H=8.62 bit/symbol, 576QAM with H=8.67 bit/symbol. The optimization results are reported in Fig. 6-left. In all cases the DAC is operated at the symbol rate adjusting the reference input clock. The highest AIR of 1.94 Tb/s is obtained with 130 GBaud DP-PCS400QAM leading also to the highest net bitrate of 1.88 Tb/s considering a known forward-error correction (FEC) code requiring 13.7 % overhead [20].

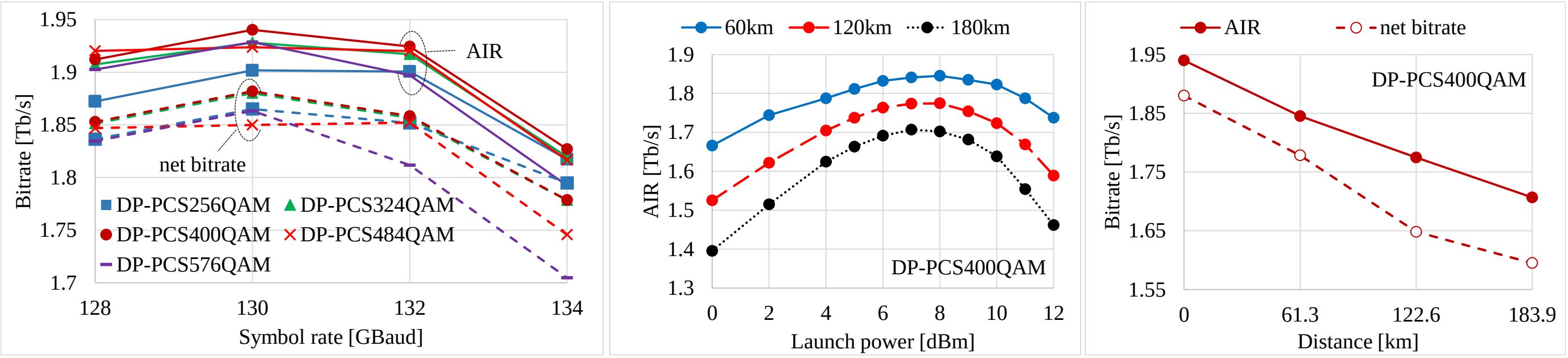

Fig. 6. Single-channel experimental results: optical B2B (left), launch power sweep (center), bitrate vs transmission distance (right).

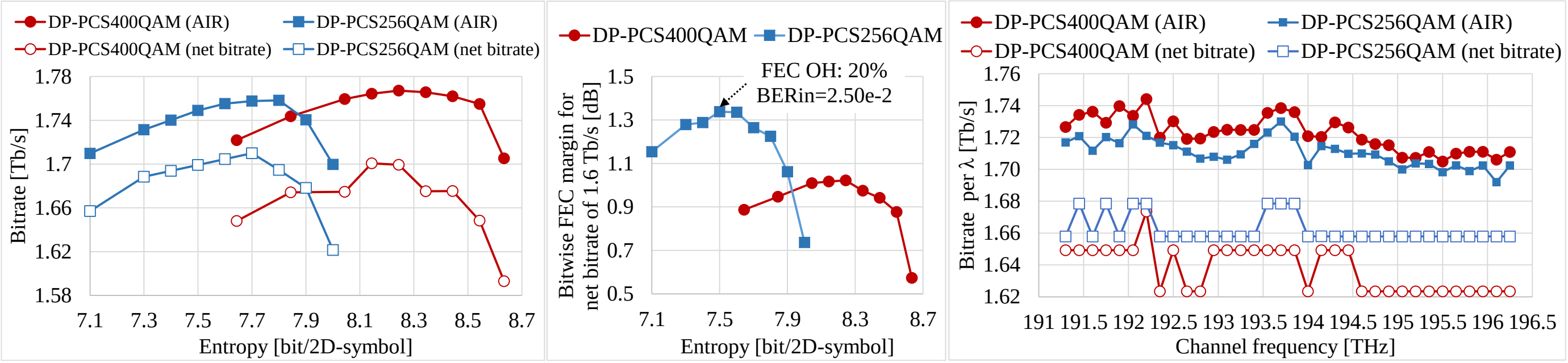

Fig. 7. Bitrate vs entropy (left) and bitwise FEC margin for net bitrate of 1.6 Tb/s vs entropy (center) for single-channel transmission, bitrate vs channel frequency (right) for 34-channel DWDM transmission over 96.5 km of field deployed ITU-T G.652 SSMF.

Fig. 6-center and Fig. 6-right show single-carrier transmission experiments performed over multiple spans of field-deployed 60 km SSMF. With 7 dBm launch power net bitrates of 1.78 Tb/s, 1.65 Tb/s and 1.60 Tb/s are achieved after 61.3 km, 122.6 km and 183.9 km transmission, respectively.

For experiments over the 96.5 km link, DP-PCS256QAM and DP-PCS400QAM have been chosen. In Fig. 7-left (single-channel) and Fig. 7-right (DWDM), we observe that DP-PCS400QAM has a higher AIR than DP-PCS256QAM. The number $m$ of bitlevels per complex symbol is 8 and 10 for 256QAM and 400QAM, respectively, since 400QAM is based on 1024QAM. The net bitrate backoff is $AIR - net_bitrate$ and the bitwise FEC margin is $(AIR - net_bitrate)/m$. Thus, considering the impact of practical FEC, following [21], the net bitrate backoff translates into a larger bitwise FEC margin for lower $m$. As we can observe in Fig. 7-center (single-channel), if we target a fixed net bitrate of 1.6 Tb/s, DP-PCS256QAM has a larger bitwise FEC margin (translated into SNR [dB]) than DP-PCS400QAM. Similarly, for a given FEC margin, a lower $m$ results in a smaller net bitrate backoff. For this reason, we observe in Fig. 7-left and Fig. 7-right, that when selecting from a set of practical codes with similar FEC margins [20], DP-PCS256QAM achieves a higher net bitrate than DP-PCS400QAM. In single-channel transmission, the highest net bitrate is 1.71 Tb/s and for all 34 channels of the DWDM system the net bitrate is ≥1.66 Tb/s for DP-PCS256QAM and above 1.62 Tb/s for DP-PCS400QAM. Fig. 8 shows the stability of the transmission system monitoring the AIR for the worst DWDM channel over the course of 14 hours, with measurements taken every 45 seconds.

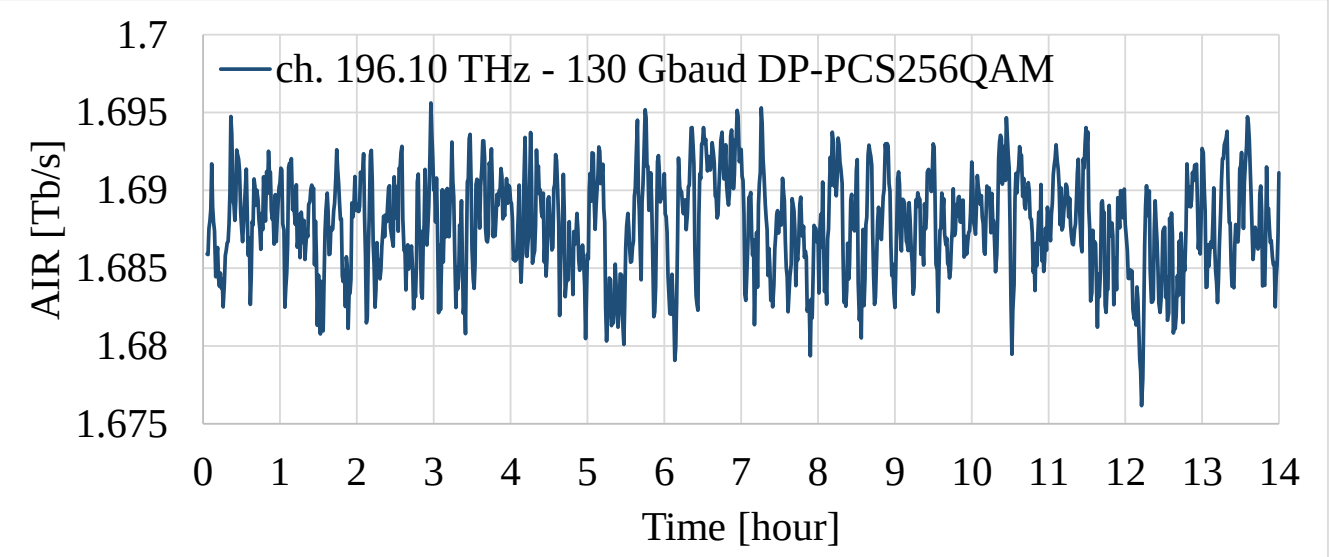

Fig. 8. AIR monitoring over 14 hours for the worst DWDM measured channel at frequency 196.10 THz.

## IV. Conclusions

We have reported record symbol rates up to 134 GBaud using one DAC per signal dimension, record net bitrates up to 1.88 Tb/s employing high-order modulation formats up to DP-PCS576QAM in optical B2B and record single-channel transmission achieving net bitrates ≥1.6 Tb/s up to 183.9 km of field-deployed fiber using 130 GBaud DP-PCS400QAM. In addition, we reported a capacity record of 56.51 Tb/s in a 34-channel DWDM configuration over a 96.5 km link with 23 dB attenuation. All 34 carriers achieve error-free net bitrates ≥1.66 Tb/s using DP-PCS256QAM. Transmission experiments were performed on a field-deployed fiber network infrastructure of Deutsche Telekom in Berlin, Germany, using EDFA-only amplification of the C-band in ITU-T G.652 SSMF.